\documentclass{article}
\usepackage[utf8]{inputenc}
\usepackage{authblk}
\usepackage{hyperref}
\usepackage{bm}
\usepackage{amsmath}

\title {Convergences and Divergences: Einstein, Poincaré, and Special Relativity}
\author{Galina Weinstein\thanks{The Department of Philosophy, University of Haifa.}}

\begin{document}

\maketitle

\begin{abstract}
Jean-Marc Ginoux’s recent book, \emph{Poincaré, Einstein and the Discovery of Special Relativity: An End to the Controversy} (2024), seeks to close the debate over the respective roles of Poincaré and Einstein. Yet what is presented as an “end” may instead invite a more careful analysis of how similar equations can conceal divergent conceptions. 
The aim here is not to rehearse priority disputes but to show how Einstein’s ether-free, principle-based kinematics marked out a path that, unlike its contemporaries, became the canonical form of special relativity. To this end, I reconstruct side by side the 1905 derivations of Poincaré and Einstein, tracing their similarities and, more importantly, their differences.
This paper reconstructs, in a novel way, the 1905 derivations of Einstein and Poincaré, highlighting their contrasting paths.  
\end{abstract}

\section{Introduction}

Jean-Marc Ginoux’s recent book, \emph{Poincaré, Einstein and the Discovery of Special Relativity: An End to the Controversy} (2024), seeks to close the debate over the respective roles of Poincaré and Einstein. Yet what is presented as an “end” may instead invite a more careful analysis of how similar equations can conceal divergent conceptions. 

The aim here is not to rehearse priority disputes but to show how Einstein’s ether-free, principle-based kinematics marked out a path that, unlike its contemporaries, became the canonical form of special relativity. To this end, I reconstruct side by side the 1905 derivations of Poincaré and Einstein, tracing their similarities and, more importantly, their differences.

This paper reconstructs, in a novel way, the 1905 derivations of Einstein and Poincaré, highlighting their contrasting paths.  

\section{Between Convention and Innovation}

\subsection{The Elephant in the Room}\label{PT}

Ginoux’s recent book reopens the long-standing Poincaré–Einstein priority discussion, interpreting “priority” in the sense of chronological precedence. He attributes to Poincaré, before June 1905, explicit possession of key mathematical structures—the Lorentz transformation’s group property, the relativistic velocity-addition law, and an articulated “principle of relativity”—together with earlier analyses of local time \cite{Gin}. From this formalist standpoint, Einstein’s 1905 paper can be read primarily as a kinematic restatement of the apparatus Poincaré had already assembled. If “theory” is defined chiefly by invariant equations, such an account naturally stresses formal overlap while downplaying conceptual discontinuity.

Ginoux also comments on Einstein’s 1955 letter to Carl Seelig, in which Einstein professed ignorance of Poincaré’s 1905 note \cite{Poi05-1} and earlier Lorentz papers. He characterizes this as "surprising" in light of Einstein’s documented familiarity with contemporary literature, both in his published citations of the March 1905 quanta of light paper and in his role as a reviewer for the \emph{Beiblätter zu den Annalen der Physik} \cite{Gin}. 
While such claims bear on questions of influence rather than of strict priority, they invite broader historiographical reflection on the distinction between acquiring a mathematical toolkit and constructing a new conceptual architecture. 

Ginoux's tone remains formally restrained, yet the implications for Einstein’s independence are left for the reader to weigh. While Ginoux refrains from alleging direct appropriation, his sustained step-by-step comparisons, generally noting Poincaré’s precedence, imply both a tacit priority claim and the possibility of indirect influence. By contrasting Einstein’s "purely kinematic" treatment with Poincaré’s "dynamic" framework, Ginoux acknowledges methodological differences while maintaining that the substantive content was largely anticipated \cite{Gin}. An Einstein-centric appraisal, however, treats the recasting of shared mathematical structures into an ether-free, principle-based kinematics not as a secondary act, but as the decisive conceptual transformation constituting the theory’s inception.  

The question of Einstein’s awareness has been debated at length: did he, before mid-1905, encounter Poincaré’s mathematical results—Lorentz transformations, group properties, relativistic velocity-addition law—and, if so, did these shape his own formulation? Contemporary exchanges shed some light. In February 1955, responding to Seelig’s inquiry, Einstein stated that "There is no doubt that the special theory of relativity, if we regard its development in retrospect, was ripe for discovery in 1905. Lorentz had already observed that for the analysis of Maxwell's equations the transformations which were later known by his name are essential, and Poincaré had even penetrated deeper into these connections." However, he stressed that he knew Lorentz’s 1895 work but neither Lorentz’s later writings nor Poincaré’s related investigations, and that his 1905 work was “in this sense” independent. What was new, he emphasized, was recognizing that the Lorentz transformation applied beyond electrodynamics, reaching the general structure of space and time, and that Lorentz invariance was a universal constraint on physical theories. This reply, published by Seelig in May 1955 \cite{Born1, Born2}, was prompted by Seelig’s reading of Edmund Whittaker’s \emph{History of the Theories of Aether and Electricity}, which had attributed the “relativity of Poincaré and Lorentz” \cite{Whi53} with foundational status. Historical evidence shows Einstein reading Lorentz’s \textit{Versuch} \cite{Lorentz1895} by 1902 \cite{CPAE1}, Doc. 130.

Einstein underlined “special” in his handwritten letter to Seelig, marking the scope of his remarks, and reaffirmed that he was unaware of Poincaré’s dynamical studies before 1905 (EA 39-070). Max Born, speaking in July 1955 at the Bern relativity conference, quoted from the published reply—omitting the underlining—thereby introducing the exchange into the public record. Einstein’s death in April left the “hot” questions unanswered; the resulting historiographical gap has ensured that debate over influence and priority continues.

Einstein’s 1905 relativity paper could have been framed in the citation-rich style of the light-quantum paper, acknowledging predecessors explicitly. Instead, it adopted a deliberately self-contained presentation, deriving the theory “from first principles” without appeal to prior results. This rhetorical strategy reinforced its conceptual autonomy but left room for later readings, such as Ginoux’s, that construe the absence of citation as significant. The irony is that the very independence of presentation—essential to Einstein’s principle-based approach—also made it easier to interpret the work as continuous with, rather than conceptually distinct from, Poincaré’s.

In 1906, Einstein acknowledged that certain formal considerations he had employed in connection with mass–energy conservation had appeared in the Poincaré 1900 paper \cite{Poi00}, while adding that he had refrained from using them directly “for the sake of clarity” \cite{Einstein06}. The documentary record establishes that by 1906, he had read Poincaré’s 1900 paper, but leaves unresolved whether this occurred before or only after the formulation of special relativity in 1905.  

This uncertainty has sustained divergent readings: for those inclined toward a Poincaré-centered account, an earlier date offers a natural line of influence; for those emphasizing Einstein’s independence, a later date aligns with the evidence. Paradoxically, the very existence of such partial documentation does not settle the question but instead preserves the ambiguity. In this respect, the episode acquires a certain “relativistic” quality of its own: the ordering of events appears to shift with the observer’s frame, and even when a connection can be established, its temporal coordinate remains flexible enough to support more than one trajectory.

In his reassessment of the formative years of special relativity, Gerald Holton engages with the view—given its most influential expression by Whittaker—that Einstein’s 1905 formulation substantially depended on earlier work by Poincaré. Holton notes that no documentary evidence links Einstein to Poincaré’s 1904 or 1905 publications before the submission of his paper, and that nothing in Einstein’s text reflects reliance on Poincaré’s conceptual framework. The omission of citations to either Poincaré or Lorentz’s 1904 paper, he suggests, is best read in light of Einstein’s normal practice of acknowledging sources he actively used; in the very same paper, Einstein twice names Lorentz when referring to the electron theory as presented in the 1895 \emph{Versuch}, which he had read \cite{Einstein05, Holton1960}.

Holton situates Einstein’s work within a vibrant research climate: between 1902 and 1905, eight papers appeared in the \emph{Annalen der Physik} on the electrodynamics of moving bodies, underscoring the field’s centrality. In this setting, Einstein’s contribution appears as the continuation of a live, multi-authored line of inquiry rather than a solitary rupture. Nevertheless, Holton underlines that Einstein did not employ Lorentz’s 1904 transformations. Even without relying on Einstein’s retrospective statements, he identifies four internal features of the 1905 paper that, in his view, point to independence from Lorentz’s 1904 formulation. Where Lorentz (and Poincaré) began from the transformations as a given, Einstein deduced them from two postulates—the relativity principle and the constancy of the speed of light—thus arriving by a distinct route \cite{Holton1960}.%
\footnote{A detailed derivation is provided in my recent paper \cite{Weinstein}.}

Holton also contrasts the formal reach of the theories. Lorentz’s 1904 scheme was restricted to small $v/c$ because of a velocity-dependent constant $\ell$, while Poincaré set $\ell = 1$, obtaining a transformation valid up to $c$. This extension, however, retained elements such as stabilizing stresses for the electron and the ether hypothesis—devices Einstein eliminated in favor of a principle-based framework. In Poincaré’s presentation, simultaneity, time measurement, and the operational meaning of coordinates remained within the conceptual boundaries of the ether theory.

By recognizing both the prominence of the topic and Poincaré’s mathematical contributions, Holton grants him a significant place in the prehistory of special relativity. Yet his reading—based on textual practice and internal analysis—presents Einstein's 1905 formulation as a decisive reconceptualization. The removal of auxiliary scaffolding and the relocation of the Lorentz transformation into an ether-free, principle-based kinematics was achieved by Einstein alone \cite{Holton1960}.

Poincaré’s \emph{Science and Hypothesis} \cite{Poi02} contains his 1898 account of clock synchronization by telegraphic signals \cite{Poi98} and a mechanically framed principle of relativity. Einstein almost certainly read and discussed the book with the Academia Olympia \cite{ES}, and parallels between its content and his 1905 operational definitions have been noted. Yet by 1905, Einstein had developed independent routes to both synchronization and a relativity principle, each pointing toward a kinematics in which simultaneity is constitutive and the constancy of light is postulated outright.%
\footnote{See the full derivation in my recent paper \cite{Weinstein}.}
Taking his 1955 statement at face value, the special theory of relativity, as he formulated it in 1905, was conceived without knowledge of Poincaré’s work on electron dynamics.

In the end, the decisive issue is not merely who arrived first, but who carried the correct clocks. 

\subsection{Between Bern and Paris: No Telegraph}

The extant record offers no direct indication that Poincaré’s May–June 1905 corrections and innovations (\cite{Wal}, Letters to Lorentz, 38.3--38.5; \cite{Poi05-1}) reached Einstein before the completion of \emph{On the Electrodynamics of Moving Bodies} \cite{Einstein05}. At that time, the 25-year-old patent examiner in Bern stood outside the scholarly correspondence networks through which such material typically circulated and had not yet met Lorentz. Surviving documentation records no communication from Lorentz to Einstein in this period, and Einstein’s first known exchange with a leading academic—his correspondence with Max Planck—dates from roughly a year later. The mid-May 1905 Poincaré–Lorentz letters (\cite{Wal}, Letters to Lorentz, 38.3--38.5) establish the state of Poincaré’s work but provide no evidence of Einstein's access to them. Nor is there a record of his consulting Poincaré’s brief \emph{Comptes rendus} note of 5 June 1905 \cite{Poi05-1} before submitting his own paper on 30 June. In the absence of corroborating materials—diaries, correspondence, library slips, marginalia, or contemporary testimony—linking Einstein to these sources in that narrow window, any claim of direct intellectual transmission rests on inference rather than documented fact.

By mid-May 1905, Einstein was already describing to Conrad Habicht a relativity paper “in progress” \cite[Doc. 28]{CPAE5}, and discussing its kinematical content with acquaintances such as Joseph Sauter, Michele Besso, and Lucian Chavan \cite{Fluckiger1974}. According to Sauter’s later memoir, Einstein had already formulated the light-signal synchronization procedure and the operational definition of simultaneity \cite{Sauter1960}. These attestations place the core kinematical construct in Einstein’s hands before any plausible opportunity to consult Poincaré’s May–June writings. 

Peter Galison’s book \emph{Einstein’s Clocks, Poincaré’s Maps} argues that Einstein and Poincaré developed overlapping but independent approaches to simultaneity, each shaped by their local technological and institutional environments \cite{Gal}.
As a leading figure at the French Bureau of Longitude, Poincaré confronted the practical challenges of global synchronization: telegraph delays, signal corrections, and human factors. His philosophy of conventions grew out of this milieu. Simultaneity and temporal standards are not metaphysical truths, but negotiated agreements chosen for convenience, continuity, and practical utility. For Poincaré, telegraphy was not a metaphor but a material model for thinking about time. He interpreted Lorentz’s “local time” in terms of synchronized telegraphic clocks, but still held onto the concept of the ether, aiming to reconcile new practices with classical frameworks.

In Bern, Einstein lived amid Switzerland’s electrochronometric culture, characterized by public clock standardization, relay-synchronization systems, and a patent office filled with applications for time-coordination devices. This environment, combined with his training at the ETH (with its pragmatic, engineering orientation), encouraged an operational approach: time is what clocks measure, and light-signal exchanges define simultaneity. In 1905, Einstein reframed simultaneity as a constitutive procedure, not a convention. He abolished the ether and elevated the invariance of $c$ to a postulate, building relativity as a principle theory of kinematics.

Galison presents Einstein and Poincaré as inhabiting parallel but nonintersecting universes. Both grappled with the same cultural pressures—railroads, telegraphy, standardized time—but from different institutional vantage points. Poincaré, the established savant, wove conventions into existing frameworks; Einstein, the outsider, used synchronization to redefine the very meaning of time \cite{Gal}.

Poincaré’s synchronization and Einstein’s synchronization may look similar at the procedural level, but they are embedded in fundamentally different conceptual frameworks. For Poincaré, synchronization is a convention, a convenient fiction layered over an underlying “true time” defined by motion through the ether. The procedure yields local time, an "apparent" valuable time for observers in uniform motion, but not the fundamental time. Poincaré still assumed the ether. The procedure only “works” because observers are ignorant of their absolute motion through it. Synchronization is a pragmatic patch, not a reconstruction of physics. Poincaré considers synchronization to be an adjustment rule in the service of celestial mechanics and longitude measurements — a technical fix that makes dynamics manageable.

For Einstein, synchronization is not a convention masking reality — it is what time means. There is no hidden “true” time. By declaring that the ether is superfluous, Einstein promoted the synchronization procedure from a convention to the operational definition of simultaneity. There is no “true” time to compare against. The light-signal procedure is definitive, not provisional. The relativity of simultaneity emerges not as a correction but as the core of the new kinematics. 

Even if Einstein had known Poincaré’s 1898 paper, the conceptual leap from “conventional synchronization under the ether” to “synchronization as the very definition of time, with no ether” is so radical that appropriation does not explain it. Thus, Einstein did not need Poincaré’s 1898 or 1900 procedure. The procedure was in the air: railway clocks, telegraphy, and astronomical longitude. Everyone in scientific Europe was aware of signal-based synchronization. What Einstein did was to invert its meaning: what for Poincaré was a practical fiction became, for Einstein, the essence of time itself.

\subsection{Priority Thread, in One Breath}

Ginoux’s book adopts a formalist, sequence-oriented historiography, in which the systematic collation of equations, dates, and correspondence is used to reconstruct the relative timing and scope of contributions. On this basis, he attributes to Poincaré, by May–June 1905, a body of results encompassing the corrections to Lorentz’s 1904 formulas, the symmetric transformation form with $\ell = 1$, the group property, and the relativistic velocity–addition law as presented in the June 5 \emph{Comptes rendus} note \cite{Gin}. For Ginoux, these achievements, combined with Poincaré’s articulation of the relativity principle, constitute the formal underpinnings of special relativity.  

In adopting what might be termed a "record-straightening" approach, Ginoux secures Poincaré’s formal priority with considerable precision. At the same time, such a framing necessarily foregrounds formal correspondences over programmatic distinctions. The operational redefinition of time and simultaneity, as well as the elimination of the ether—central to historians such as John Stachel \cite{Stachel2002}—receive proportionally less emphasis.  

Ginoux structures his account of Einstein’s 1905 paper around a closely sequenced chronology, drawing attention to the temporal proximity of Poincaré’s publications and public lectures to Einstein’s submission date. Verbal correspondences, such as the parallel between the title of Einstein’s paper and a phrase from Poincaré’s 1904 Saint Louis lecture \cite{Poi4}, are noted alongside recurrent juxtapositions of formulations that are often described as "identical" in substance, despite differences in expression. The cumulative effect is to invite the inference that these parallels are unlikely to be accidental.  

A series of technical correspondences is cataloged—Lorentz transformations, the relativistic velocity–addition law, the Doppler effect, and the invariance of Maxwell’s equations—each of which appears in Poincaré’s work shortly before Einstein’s \cite{Gin}. 

\subsection{The Ghost Prefactor}

Particular emphasis falls on Einstein’s introduction of a scale factor:
\begin{equation} \label{Pos}
\phi(v)\;=\;a(v)\,\frac{1}{\sqrt{1-\frac{v^2}{c^2}}},    
\end{equation}
into the Lorentz transformation. 

It is essential to identify the role of undetermined functions in Einstein’s method and the stepwise operational logic of the derivation of the Lorentz transformation in his 1905 relativity paper:%
\footnote{A detailed derivation is provided in my recent paper \cite{Weinstein}.}

\begin{enumerate}
  \item Einstein begins from the operational definition of simultaneity in $k$:
  \begin{equation} \label{30}
  \frac{\tau_0+\tau_2}{2}=\tau_1,    
  \end{equation} 
  combined with the light postulate, this yields a \emph{functional equation} for $\tau(x',t)$ (with $x'=x-vt$ used as a Galilean placeholder).

  \item A first–order expansion of the synchronization condition \eqref{30} leads to the partial differential equation (PDE):
  \begin{equation} \label{65}
   \frac{\partial \tau}{\partial x'}+\frac{v}{c^2-v^2}\,\frac{\partial \tau}{\partial t}=0,\qquad 
    \frac{\partial \tau}{\partial y}=0,\quad \frac{\partial \tau}{\partial z}=0,   
  \end{equation}
    This already constrains $\tau$ to depend on a \emph{specific} linear combination of $x'$ and $t$. 

  \item The general solution of the PDE is:
  \begin{equation} \label{75}
   \tau(x',t)=F\!\left(t-\frac{v}{c^2-v^2}\,x'\right),  
  \end{equation}
  with $F$ arbitrary. Imposing linearity yields:
  \begin{equation} \label{76}
  F(T)=a(v)\,T.    
  \end{equation}
  
  \item Enforcing the light postulate in multiple directions produces the provisional linear transformation:
  \begin{equation}\label{eq52}
  \begin{aligned}
   &\xi \;=\; a(v)\,\gamma^{2}\,\bigl(x - v t\bigr), 
   \quad
   \tau \;=\; a(v)\,\gamma^{2}\!\left(t - \frac{v}{c^{2}}\,x\right), \\
   &\eta \;=\; a(v)\,\gamma\,y, \quad 
   \zeta \;=\; a(v)\,\gamma\,z.
   \end{aligned}
   \end{equation} 
    At this stage, the scale factor $a(v)$ remains undetermined. 
  
   \item  Substituting the reparametrization \eqref{Pos}:
    
  \begin{equation} \label{77}
  \phi(v)=a(v)\,\gamma,    
  \end{equation} 
  into the transformation \eqref{eq52}, yields the linear transformation:

  \begin{equation} \label{Pr}
  \tau=\phi(v)\,\gamma\!\left(t-\frac{v}{c^2}x\right),\quad 
    \xi=\phi(v)\,\gamma\,(x-vt),\quad 
    \eta=\phi(v)\,y,\quad 
    \zeta=\phi(v)\,z.    
  \end{equation}
  
  \item Only after introducing the most general linear form consistent with the postulates does Einstein apply reciprocity and transverse symmetry (together with continuity at $v=0$. These physical requirements then \emph{fix} the free factor to:
  
  \begin{equation} \label{40}
   \phi(v)=1.   
  \end{equation}
  
  This conclusion stems from the principle of relativity and isotropy.
\end{enumerate}

Ginoux, echoing Arthur I. Miller’s remark that “it seems as if [Einstein] knew beforehand the correct form of the set of relativistic transformations” [\eqref{Pr} and \eqref{eq12}] \cite{Miller}, treats the substitution [\eqref{Pos} or \eqref{77}] as a justificatory gap—indeed a hint of foreknowledge—emphasizing that Einstein cites neither Lorentz nor any other source \cite{Gin}. In short, Ginoux claims that Einstein “knew beforehand” the Lorentz form and thus defined $\phi(v)=a(v)\gamma$ \eqref{77}.

Ginoux treats \eqref{Pos} as a purposeful nudge toward the Lorentz transformation rather than a neutral reparametrization. 
Therefore, he invokes Poincaré’s route (see section \ref{dyn}). He asserts that from the transformation (7.21) [i.e., \eqref{eq52}] one can “easily” recover the relativistic velocity addition law already obtained by Poincaré in his \emph{Rendiconti} memoir and C.R.A.S. note \cite{Gin}, thereby gesturing toward Poincaré’s velocity transformation route \cite{Poi05-2}.

Let us indulge Ginoux’s cue and momentarily recast Poincaré’s velocity transformation in Einstein’s rods-and-clocks notation—if only to watch the undetermined scale slip quietly away—mindful that Einstein himself never took this route. 

Let there be a material point whose position at time $t$ in the system $K$ is given by the coordinates $x,y,z$. In the interval of time $dt$ this point moves through the distances $dx,dy,dz$. We therefore call:

\begin{equation}
u_x=\frac{dx}{dt},\qquad u_y=\frac{dy}{dt}\qquad u_z=\frac{dz}{dt}    
\end{equation}
the velocity components of the point measured in $K$.

According to the transformation equations \eqref{eq52}, we have for the differentials of the coordinates in $k$:

\begin{equation}
d\xi = a(v)\,\gamma^{2}\,[dx - v\,dt], 
\qquad
d\tau = a(v)\,\gamma^{2}\!\left[dt - \frac{v}{c^{2}}\,dx\right].
\end{equation}
Hence, the component of the velocity in the direction of the $x$-axis, as measured in $k$, is:

\begin{equation}
u'_x\equiv\frac{d\xi}{d\tau}
 \;=\; \frac{a(v)\gamma^{2}\,[dx - v\,dt]}{a(v)\gamma^{2}\!\left[dt - \frac{v}{c^{2}}\,dx\right]}
 \;=\; \frac{a(v)\gamma^{2}\,[\frac{dx}{dt} - v\,\frac{dt}{dt}]}{a(v)\gamma^{2}\!\left[\frac{dt}{dt} - \frac{v}{c^{2}}\,\frac{dx}{dt}\right]}
 \;=\; \frac{u_x-v}{1-\frac{u_x v}{c^{2}}}\,.  
\end{equation}
For the transverse coordinates, we obtain from the transformation \eqref{eq52} $d\eta = a(v)\,\gamma\,dy, \, d\zeta = a(v)\,\gamma\,dz $, so that the corresponding velocity component in $k$ is:
\begin{equation}
u'_y\equiv\frac{d\eta}{d\tau}
 \;=\; \frac{a(v)\gamma\,dy}{a(v)\gamma^{2}\!\left[dt - \frac{v}{c^{2}}\,dx\right]}
 \;=\; \frac{u_y}{\gamma\,\Bigl(1-\frac{u_xv}{c^{2}}\Bigr)}.  
\end{equation}
And similarly for $\zeta$:
\begin{equation}
u'_z\equiv\frac{d\zeta}{d\tau}
\;=\; \frac{a(v)\gamma\,dz}{a(v)\gamma^{2}\!\left[dt - \frac{v}{c^{2}}\,dx\right]}
\;=\; \frac{u_z}{\gamma\,\Bigl(1-\frac{u_xv}{c^{2}}\Bigr)}.    
\end{equation}

Ginoux's argues: 

1) Already at the provisional stage [equation \eqref{eq52}], if we compute velocity transformations, the undetermined $a(v)$ cancels. That is, the relativistic velocity addition law (which Poincaré had published in the \emph{Rendiconti} paper) drops out even without fixing $a(v)$. The common factor $a(v)\gamma^{2}$ cancels identically in $u'_x$, and the ratio $a(v)\gamma/[a(v)\gamma^{2}]$ reduces to $1/\gamma$ in $u'_{\perp}$. 

2) Therefore, he suggests that Einstein’s redefinition \eqref{Pos} was not a neutral mathematical convenience but a way of nudging the transformation toward the Lorentz form he already knew. Ginoux infers: "...we come to the conclusion that Einstein has necessarily posed \eqref{Pos} \cite{Gin}. 

In other words, Ginoux argues that Einstein quietly absorbed a $\gamma$ factor into $a(v)$ because he had the Lorentz transformation in mind, without justifying it from first principles or citing predecessors (Lorentz, Poincaré, Woldemar Voigt) \cite{Gin}.

In truth, Einstein’s procedure was altogether more modest. Instead of presupposing the final form, he allowed the equations to speak for themselves, and only afterward fixed the undetermined factor in keeping with the principle of relativity. The cancellation of $a(v)$ in the velocity transformation shows only that the velocity-addition formula does not depend on the scale factor. It does not show that Einstein could have left $a(v)$ arbitrary and still had a fully consistent kinematics.
Physical consistency (reciprocity, isotropy, invariance of $c$) forces introducing the combination \eqref{Pos} and then setting $\phi(v)=1$ \eqref{40}, which yields the Lorentz transformation \eqref{eq12}. That is what Einstein actually argued in his paper.

Einstein did not use Poincaré’s route through the velocity transformation; he did \emph{not} derive the relativistic velocity addition law from \eqref{eq52} while the overall scale was undetermined. Using the final Lorentz transformation \eqref{eq12}, after fixing $a(v)$ and $\phi(v)=1$, he obtained the relativistic addition law \cite{Einstein05}:
\begin{equation} \label{eq41}
u \;=\; \frac{v + w}{1 + \tfrac{vw}{c^{2}}}\,.
\end{equation}

So, Ginoux’s assertion that Einstein “necessarily posed” equation \eqref{Pos} effectively suggests that Einstein already knew the final Lorentz form and deliberately tailored his definitions to reproduce it. Yet this interpretation projects hindsight onto Einstein’s 1905 reasoning that is not supported by the structure or style of his original argument.

If Einstein had imported Lorentz’s result, he would not have left $a(v)$ undetermined through much of the derivation, nor would he have carried the arbitrary factor explicitly in the provisional equations \eqref{Pr}. Instead, the sequence:
\begin{equation}
\begin{aligned}   
&\text{sync condition} \eqref{30}\;\Rightarrow\;
\text{PDE } \eqref{65}\;\Rightarrow\;
\text{general solution } \eqref{75} \\
&\Rightarrow\;
\text{linearity } \eqref{76}\;\Rightarrow\;
\text{provisional transformation } \eqref{eq52} \\ 
&\;\Rightarrow\; 
\text{provisional Lorentz transformation with }\phi(v)\text{ in } \eqref{Pr} \\
&\;\Rightarrow\;
\text{reciprocity/symmetry }\Rightarrow\;
\phi(v)=1\text{ in } \eqref{40},
\end{aligned}
\end{equation}
shows a \emph{stepwise, operational} derivation grounded in Einstein’s postulates. The reparametrization in \eqref{77} is a notational convenience, not a tacit assumption of the final Lorentz form. The factor $\phi(v)$ is fixed \emph{by the relativity principle and isotropy}, demonstrating that Einstein was not borrowing a result from Lorentz’s electron theory but re-deriving the transformation within a new conceptual framework.

Ginux suggests that the paucity of citations in 1905 indicates dependence on Lorentz/Poincaré. Regardless of historical editorial practice, that claim is orthogonal to the logic of the derivation. The presence of the free factor $\phi(v)$ throughout, and its elimination by reciprocity/isotropy at the end, is incompatible with beginning from the Lorentz form. The mathematics of \eqref{30}, \eqref{65}, \eqref{75}, \eqref{76}, \eqref{Pr}, and \eqref{40} already displays an \emph{independent, operational} derivation.

In short, when reconstructed carefully, Einstein’s method exhibits a progression from synchronization to the Lorentz transformation that is incompatible with the “retrofitting” hypothesis. The presence and retention of undetermined functions until the physical constraints are applied is precisely what one expects from an \emph{independent} derivation driven by the postulates themselves. 

\subsection{Poincaré's Ether vs. Einstein's Ether}

Ginoux presses the point that the familiar contrast—Poincaré and Lorentz clinging to the ether while Einstein banished it—rests on fragile ground. He reminds the reader that Einstein himself, once gravity entered the picture, spoke of a “gravitational ether” or “Mach ether.” From this perspective, the divide dissolves. Rather than two theories separated by the presence or absence of ether, there is a single edifice articulated in different idioms. As Ginoux notes, from 1916 onward, Einstein reintroduced the term “ether” into the framework of general relativity. He argues that this shows Einstein, like Poincaré, required what Emil Cohn had called a “heuristic concept,” or in Poincaré’s phrase, a “convenient hypothesis,” to render the new theory intelligible. In the same spirit, Ginoux revisits Einstein’s 1905 definition of simultaneity: “a clock at rest in a system at rest.” Paraphrasing Poincaré’s objection—if there is no absolute space, with respect to what is this clock at rest?—he suggests that Einstein’s framework itself presupposes a resting background. The result, Ginoux concludes, is a symmetry. Neither Einstein nor Poincaré wholly abolished the ether; they only redescribed its role \cite{Gin}.

In Einstein’s 1905 paper, however, the “rest system” $K$ is a fiducial inertial frame introduced for expository convenience, possessing no privileged ontological status among the continuum of inertial frames. Where Poincaré clothed invariance in the language of group theory and ether, Einstein reduced it to rods, clocks, and operational prescriptions. The ether did not so much vanish as become irrelevant.

When Einstein did speak of an ether, it was in a sense that differed fundamentally from Lorentz’s construct. The Leiden inaugural lecture of October 27, 1920—delayed by the formalities of royal appointment—was a case in point. Paul Ehrenfest had once jokingly proposed the Dutch title “Down with the ether superstition!!” Still, when Einstein finally took the stage, he did not so much banish the ether as redefine it \cite{CPAE9}, Doc.~373. He described a “new ether” or “world-matter,” a Machian medium required as a carrier of inertial effects in a universe where action at a distance was no longer admissible. This was not Lorentz’s rigid, absolute frame. Still, the metric field $g_{\mu\nu}$ itself—varying from place to place, determined by material phenomena, and never assignable to an independent state of motion \cite{Wei1}.

It is sometimes repeated, with a brevity that conceals more than it reveals, that “the ether he reintroduced differed fundamentally from the ether he had banished” \cite{Stachel2001}. The phrase is not wrong, but it hardly captures the mathematics of general relativity, where the metric tensor is no ether at all. To read it otherwise is to replace analysis with a slogan.\footnote{A fuller account of Einstein’s path to this insight, and of why the metric cannot be reduced to an “ether,” is given in my study \cite{Wei3}.}

In a 1916 letter to Lorentz, Einstein himself, half-diplomatically and half-seriously, had remarked that general relativity was “closer to the ether hypothesis” than special relativity \cite{CPAE8}, Doc. 222. The remark was a gesture of respect: Lorentz still clung to ether, and Einstein, who revered him, framed his own theory in language that Lorentz could recognize. The Leiden lecture was thus homage as much as physics—a theatrical bow to Lorentz, even as the stage and script had already shifted.

\subsection{The Light Postulate}

Ginoux claims that Poincaré’s adoption of $c = 1$ in his equations stemmed from the experimental result of Michelson and Morley \cite{MM}. In contrast, for Einstein, the invariance of the velocity of light was a necessary postulate for establishing the Lorentz transformation. Ginoux regards Einstein’s 1905 light postulate as “curious,” since it appears to place the invariance of $c$ at the origin of the relativity principle, rather than the other way around. He invokes Jean-Marc Lévy-Leblond’s later demonstration that the Lorentz transformation can be derived from the relativity principle and general symmetry assumptions alone, without assuming the constancy of the speed of light, as if to suggest that Einstein elevated the light postulate to an unnecessarily prominent status \cite{Gin}. 

In Lévy-Leblond's construction, the relativity principle, together with spatial isotropy, homogeneity, and reciprocity, all produce a 
one-parameter family of transformation groups characterized by an invariant speed $\alpha$, whose value must be fixed empirically. In our world, electrodynamics gives $\alpha = 1/c^{2}$. Lévy-Leblond’s analysis thus dispenses with the light postulate as an \emph{axiom}, but does not make it a \emph{deduction} from the relativity 
principle alone. The identification of $c$ as the limiting speed remains an essential, physically motivated step — one that Einstein’s 1905 formulation incorporates from the outset with full conceptual economy.

In Lévy-Leblond’s notation, the most general inertial transformation consistent with homogeneity, isotropy, reciprocity, and the group property has the form 
\cite{LL}:
\begin{equation} \label{eq17}
x' = \lambda(v)\,\gamma(v)\,(x - v t), \qquad
t' = \gamma(v)\,(t - \alpha(v) x),
\end{equation}
with $\lambda(v)$ and $\alpha$ to be determined. These conditions, together with a mild causality requirement, yield: $\lambda(v)=1, \quad \gamma(v)=(1-\alpha v^{2})^{-1/2}$. The constraint $\lambda(v)=1$ is precisely Poincaré’s $\ell=1$ of 1905 (from the
group property), and Einstein’s $\varphi(v)=1$ from section \S 3 of his 1905 relativity paper (derived from reciprocity and isotropy).

The parameter $\alpha$ in \eqref{eq17} labels three distinct kinematics: 

(1) $\alpha=0$ (Galilean),  

(2) $\alpha<0$ (ruled out by causality), and: 

(3) $\alpha>0$, where we may write $\alpha = 1/c^{2}$.

\noindent In case (3) one recovers the standard Lorentz transformation and the relativistic 
velocity--addition law:
\begin{equation}
V = \frac{v_{1} + v_{2}}{1 + v_{1}v_{2}/c^{2}},
\end{equation}
which guarantees $|V| \leq c$ for all subluminal inputs \cite{LL}.

Lévy-Leblond does not assume anything about light itself; his derivation yields only the \emph{existence} of a universal invariant speed. Identifying this $c$ with the velocity of light in vacuum requires an empirical step, supplied by electrodynamics or direct experiment. In his paper, Lévy-Leblond carried out this identification explicitly \cite{LL}. In this sense, his recourse to electrodynamics belonged to the physical interpretation, not the derivation. 

Yet Einstein’s procedure is not well captured by this inversion. In 1905, the relativity principle and the light postulate are introduced as independent, coequal assumptions: one expresses the invariance of physical law across inertial frames, the other operationalizes simultaneity by light signals. Taken together with homogeneity, isotropy, and reciprocity, they suffice to yield the Lorentz transformation with $\varphi(v)=1$.

Lévy-Leblond’s reconstruction is indeed elegant, but it leaves the invariant speed as an undetermined constant, $\alpha^{-1/2}$, to be identified empirically. Einstein’s step was precisely to identify this constant with the speed of light from the outset, motivated by a web of physical tensions—from ether-drift nulls to the chasing-a-light-beam paradox—that pointed directly to light as the carrier of invariance. What Ginoux calls “curious” was in fact Einstein’s simplicity. By letting $c$ serve both as the synchronization signal and the invariant limit, the contradictions between mechanics and electrodynamics collapsed into a single principle-based framework.
From a group-theoretic standpoint, Lévy-Leblond shows that the light postulate is not strictly logically necessary. From Einstein’s perspective in 1905, it was physically indispensable.

\section{Einstein's and Poincaré’s Derivations}

\subsection{Poincaré's May 1905 Letters to Lorentz} \label{W}

In his May 1905 letter to Lorentz, Poincaré wrote that 
Lorentz had proposed that the transformed quantities obey \cite{Wal}, letter 38.3:

\begin{equation}
\frac{1}{k l^3}\,\rho = \rho', \qquad 
k^{2} u_{x} = u'_{x}, \qquad 
k^{2} u_{y} = u'_{y},    
\end{equation}
but he pointed out that these relations violated charge conservation. 
To remedy this, he introduced a corrective factor, suggesting instead (Poincaré Archives Nancy):

\begin{equation} \label{G-1}
\frac{1}{kl^3}\,{\rlap{/}(}\,\rho(1+\varepsilon v_x) = \rho'\qquad 
\frac{1}{kl^3}\,{\rlap{/}\rho}\,(v_x+\varepsilon) = \rho' u_x' 
\end{equation}
This adjustment ensured compatibility with the continuity condition. 

In his letter to Lorentz, Poincaré did not derive the full set of transformations for charge density and current. However, in the \emph{Comptes rendus} note of June 5, 1905, Poincaré writes down explicitly the relations for the transformed charge density and current components \cite{Poi05-1}:

\begin{equation} \label{eqtr} 
\rho' = \frac{k}{l^3}\rho \, (1 + \varepsilon \xi), \qquad \rho' \xi' = \frac{k}{l^3} \rho (\xi + \varepsilon), \qquad \rho' \eta' = \frac{1}{l^3}\rho \eta, \qquad \rho' \zeta' = \frac{1}{l^3}\rho \zeta. 
\end{equation} 

In his May 1905 letter to Lorentz, Poincaré proposed inserting the factor $(1 + \varepsilon v_x)$, but only in the $x$ direction, without presenting the full transformations. A month later, in his \emph{Comptes rendus} note, he wrote down the complete set of formulas \eqref{eqtr}.

Ginoux writes \cite{Gin}: "It’s also important to note that 
dividing these last two formulas [\eqref{G-1}] gives:

\begin{equation*} 
\frac{\tfrac{1}{kl^3} \rho \, (v_x + \varepsilon)}
     {\tfrac{1}{kl^3} \rho \, (1 + \varepsilon v_x)}
= \frac{\rho' v'_x}{\rho'} 
\;\;\longleftrightarrow\;\;
v'_x = \frac{v_x + \varepsilon}{1 + \varepsilon v_x}   
\end{equation*}

\vspace{2mm}
\noindent which is nothing other than the new \emph{relativistic velocity addition law}..."

In 1905, Poincaré did not explicitly derive the relativistic velocity addition law. He had the algebraic machinery in place—the transformed charge densities and factors \eqref{G-1}—but stopped short of dividing the two expressions.

At that point, Poincaré was only one algebraic step away from the velocity transformation. He had set up the correct structure, yet neither in his May 1905 letter to Lorentz nor in the 1905 \emph{Comptes rendus} note did he carry out the division that yields the velocity transformations:

\begin{equation}\label{eqr}
\xi' = \frac{\rho' \xi'}{\rho'} = \frac{\rho \xi + \varepsilon \rho}{\rho + \varepsilon \rho \xi}, \quad
\eta' = \frac{\rho' \eta'}{\rho'} = \frac{\rho \eta}{k(\rho + \varepsilon \rho \xi)}, \quad
\zeta' = \frac{\rho' \zeta'}{\rho'} = \frac{\rho \zeta}{k(\rho + \varepsilon \rho \xi)}. 
\end{equation}
Thus, in 1905, Poincaré stopped short of writing \eqref{eqr} and the velocity transformations \eqref{eq101}. From equation \eqref{eqr}, the longitudinal component is:
\begin{equation} \label{G}
\boxed{\xi' = \frac{\rho' \xi'}{\rho'} =\frac{\xi + \varepsilon}{1 + \varepsilon \xi}.}    
\end{equation}
Ginoux claims that equation \eqref{G} “is the new relativistic velocity addition law that Poincaré had already formulated in his second letter to Lorentz in May 1905” \cite{Gin}. However, this final step is absent from both the letter and the June \emph{Comptes rendus} note. Mathematically, Poincaré possessed all the elements; historiographically, what matters is that he did not actually perform or record the derivation.

In another letter to Lorentz (dated May 1905), Poincaré
writes the Lorentz transformation \cite{Wal}, letter 38.4:%
\footnote{$\varepsilon$ denotes a dimensionless velocity parameter, the velocity in units of the speed of light, set to $c=1$. The Lorentz transformation gives us relations between coordinates $(x, t, y, z)$ and $(x', t', y', z')$.}

\begin{equation} \label{eq97} 
x' = k l \, (x + \varepsilon t), \quad t' = k l \, (t + \varepsilon x), \quad y' = l y, \quad z' = l z, 
\end{equation} 
\begin{equation} 
\text{with:} \qquad k = \frac{1}{\sqrt{1 - \varepsilon^2}}, 
\end{equation} 
He applies successively two such transformations: 

\noindent First with $(k,l,\varepsilon)$, then with $(k',l',\varepsilon')$, and writes the result in the same form with new parameters $(k'',l'',\varepsilon'')$.
Comparing coefficients, he finds:
\begin{equation} \label{eq97-5}
\boxed{\varepsilon''=\frac{\varepsilon+\varepsilon'}{1+\varepsilon\varepsilon'},} 
\qquad l''=ll', 
\qquad k''l''=kk'll'(1+\varepsilon\varepsilon').
\end{equation}
For the set of transformations to form a group, one must have $l=f(\varepsilon)$ with the functional relation
\begin{equation}
f(\varepsilon'')=f(\varepsilon)f(\varepsilon').
\end{equation}
Trying $f(\varepsilon)=(1-\varepsilon^2)^m$, Poincar\'e shows that consistency forces $m=0$, hence $l=1$.
The Lorentz transformations \eqref{eq97} thus form a group with unique structure parameters, and the law \eqref{eq97-5} ensures closure of the group under composition.%
\footnote{A detailed derivation is presented in my recent paper \cite{Weinstein}.}

Ginoux states that “ Thus, by May 1905, Poincaré was already in possession of the transformation 
that leaves Maxwell’s four Eq. ... invariant, and had demonstrated that it forms an invariance group of Dynamics. This led him to establish the new relativistic velocity addition law ($\varepsilon''=\frac{\varepsilon+\varepsilon'}{1+\varepsilon\varepsilon'}$) as Miller pointed out over forty years ago" \cite{Gin}.

Why is equation \eqref{eq97-5} not the velocity addition law? The symbols $\varepsilon,\varepsilon',\varepsilon''$ are group parameters labeling Lorentz transformations (essentially, dimensionless rapidities).  
Equation \eqref{eq97-5} is therefore the \emph{group composition law} for successive boosts, not a physical law of how material particle velocities add.
The actual velocity transformations \eqref{eq101} require differentiating the Lorentz transformation \eqref{eq97} as Poincaré himself later derived in the \emph{Rendiconti di Palermo} memoir.

Historians of science sometimes read backwards into Poincaré’s or Lorentz’s writings with knowledge of Einstein’s 1905 paper, and then fish out isolated algebraic relations as if they were embryonic versions of relativity’s central concepts. But unless the author explicitly frames them as such, this risks anachronism. Poincaré did remarkable work on electrodynamics, the group properties of transformations, and he worked on the invariance of the Lorentz transformation. Still, he did not formulate relativity as a new kinematic framework. Mining his papers to extract proto–addition velocity or implicit transformations is more a reconstruction exercise than historical evidence.

\subsection{Poincaré's 1906 Derivation in the \emph{Rendiconti} paper} \label{dyn}

In the 1906 \emph{Rendiconti} paper, Poincaré begins from the continuity equation, i.e., charge conservation:  

\begin{equation} \label{eqC-2} 
\frac{\partial \rho}{\partial t} + \frac{\partial (\rho \xi)}{\partial x} + \frac{\partial (\rho \eta)}{\partial y} + \frac{\partial (\rho \zeta)}{\partial z} = 0.
\end{equation}
This frames the problem, i.e., densities and currents must transform so that continuity is maintained.

However, when computing the new velocities ($\xi', \eta', \zeta'$), he does not appeal to the continuity equation at all. 
Poincaré differentiates the Lorentz transformation \eqref{eq97} to compute the components of the velocity transformations. This argument uses only the coordinate transformation, without reference to charge density, current, or continuity. 

We start with the Lorentz transformation in the form \eqref{eq97}. 
We define the coordinate velocities in the unprimed system:
\begin{equation} \label{d}
\xi = \frac{dx}{dt}, \qquad 
\eta = \frac{dy}{dt}, \qquad 
\zeta = \frac{dz}{dt},
\end{equation}
and in the primed system:
\begin{equation}
\xi' = \frac{dx'}{dt'}, \qquad 
\eta' = \frac{dy'}{dt'}, \qquad 
\zeta' = \frac{dz'}{dt'}.
\end{equation}
Differentiating the transformation \eqref{eq97}, we obtain:
\begin{equation}
\frac{dx'}{dt} = k l \!\left( \frac{dx}{dt} + \varepsilon \right), 
\qquad 
\frac{dt'}{dt} = k l \!\left( 1 + \varepsilon \frac{dx}{dt} \right), 
\qquad 
\frac{dy'}{dt} = l \eta, 
\qquad 
\frac{dz'}{dt} = l \zeta.
\end{equation}
Substituting $\xi = \tfrac{dx}{dt}$ from \eqref{d}, this becomes:
\begin{equation}
\frac{dx'}{dt} = k l (\xi + \varepsilon), 
\qquad 
\frac{dt'}{dt} = k l (1+\varepsilon \xi),
\qquad 
\frac{dy'}{dt} = l \eta, 
\qquad 
\frac{dz'}{dt} = l \zeta.
\end{equation}
Taking ratios, we find the velocity transformations \cite{Poi05-2}:
\begin{equation} \label{1}
\xi' = \frac{dx'}{dt'} 
= \frac{kl(\xi+\varepsilon)}{kl(1+\varepsilon \xi)} 
= \frac{\xi+\varepsilon}{1+\varepsilon \xi},
\end{equation}
\begin{equation} \label{2}
\eta' = \frac{dy'}{dt'} 
= \frac{l \eta}{kl(1+\varepsilon \xi)} 
= \frac{\eta}{k(1+\varepsilon \xi)},
\end{equation}
\begin{equation} \label{3}
\zeta' = \frac{dz'}{dt'} 
= \frac{l \zeta}{kl(1+\varepsilon \xi)} 
= \frac{\zeta}{k(1+\varepsilon \xi)}.
\end{equation}
Thus, the velocity transformation laws are \cite{Poi05-2}:
\begin{equation} \label{eq101} 
\xi' = \frac{\xi + \varepsilon}{1 + \varepsilon \, \xi}, \qquad \eta' = \frac{\eta}{k \,(1 + \varepsilon \, \xi)}, \qquad \zeta' = \frac{\zeta}{k \,(1 + \varepsilon \, \xi)}. 
\end{equation}

Poincaré then plugs the velocity transformation \eqref{eq101} back into the current–density relations ($\rho', \rho' \xi', \rho' \eta', \rho' \zeta'$) \eqref{eqtr}, confirming that the transformed quantities again satisfy the continuity equation:

\begin{equation} \label{e9} 
\frac{\partial \rho'}{\partial t'} + \frac{\partial (\rho' \xi')}{\partial x'} + \frac{\partial (\rho' \eta')}{\partial y'} + \frac{\partial (\rho' \zeta')}{\partial z'} = 0. 
\end{equation}
Because $\tfrac{k}{l^{3}}$ and $\tfrac{1}{l^{3}}$ in equation \eqref{eqtr} are constants, the primed continuity equation follows immediately from these.  

In other words, the charge conservation motivates the need to transform $\rho$ and $\rho \mathbf{v}$, while the differentiation provides the algebraic form of the velocity transformations. The two are interwoven.

\subsection{Einstein's 1905 Derivation of The velocity and Charge Density Transformation} \label{VP}

Ginoux remarks: “In paragraph §9 entitled ‘Transformations of the Maxwell-Hertz equations’, Einstein rediscovers the expressions for velocity, charge, and density obtained by Poincaré as early as May 1905 (see Eq. \eqref{eqtr}, Chap. 6)” \cite{Gin}. 
Thus, according to Ginoux, by May 1905 Poincaré already held the essential foundations of the theory, weeks—if not months—ahead of Einstein, as shown by his correspondence with Lorentz. 
Ginoux’s claim warrants an explanation of Einstein's derivation of the velocity and charge transformation in section \S 9. 

\vspace{2mm}
 
First, in section \S 6, Einstein proceeds heuristically as follows \cite{Einstein05}: 
\begin{enumerate}
\item He assumes that the Maxwell--Hertz equations in empty space hold in the rest system \(K\), writing the electric and magnetic force (field) components as
$(X,Y,Z)$ and $(L,M,N)$, respectively.

\item He demonstrated in the kinematical part (in section \S 3) that for a transformation to a moving system $k$, which moves with velocity $v$ in the direction of the $x$–axis relative to the system $K$, the equations are (the Lorentz transformation):
\begin{equation} \label{eq12}
\boxed{\tau = \gamma \left( t - \frac{v}{c^{2}}x \right), \qquad
\xi = \gamma (x - v t), \qquad
\eta = y, \qquad
\zeta = z,}
\end{equation}
where $c$ is the velocity of light and: 
\begin{equation}
\boxed{\gamma = \frac{1}{\sqrt{1-\frac{v^{2}}{c^{2}}}}.}
\end{equation}

He then applies the \emph{Lorentz transformation} \eqref{eq12} from section \S3 to the space--time derivatives in the Maxwell--Hertz equations, thereby expressing these equations as they would appear in the system $k$ moving with speed $v$ along the $x$-axis relative to $K$.

\item Invoking the \emph{principle of relativity}, he requires that the Maxwell--Hertz equations in empty space hold in $k$ in the same form. Hence the fields
$(X',Y',Z')$ and $(L',M',N')$ in $k$ satisfy equations of the same functional form as in $K$.

\item Demanding that the transformed equations obtained in step (2) coincide in form with those postulated in step (3) yields the transformation laws for the electromagnetic field components.

For a boost along $x$ with $\gamma=1/\sqrt{1-v^{2}/c^{2}}$ one obtains:%
\footnote{The inverse field transformations corresponding to a boost with velocity $-v$:
\begin{equation} \label{EM-1}
\begin{aligned}
X &= X', 
&\qquad Y &= \gamma \left( Y' + \frac{v}{c}\,N' \right), 
&\qquad Z &= \gamma \left( Z' - \frac{v}{c}\,M' \right), \\[6pt]
L &= L', 
&\qquad M &= \gamma \left( M' - \tfrac{v}{c}\,Z' \right), 
&\qquad N &= \gamma \left( N' + \tfrac{v}{c}\,Y' \right).
\end{aligned}
\end{equation}}

\begin{equation} \label{EM}
\boxed{\begin{aligned}
X'&=X, &\qquad Y'&=\gamma\!\left(Y-\frac{v}{c}\,N\right), &\qquad Z'&=\gamma\!\left(Z+\frac{v}{c}\,M\right),\\
L'&=L, &\qquad M'&=\gamma\!\left(M+\frac{v}{c}\,Z\right), 
&\qquad N'&=\gamma\!\left(N-\frac{v}{c}\,Y\right).
\end{aligned}}
\end{equation}
\end{enumerate}

In section \S 9, Einstein treats the inhomogeneous Maxwell–Hertz equations with convection current
$\mathbf{j}=\rho\,\mathbf{u}$ ($\rho$ is the "density of electricity", charge density) \cite{Einstein05}: 

\begin{enumerate}
\item He begins with the Maxwell–Hertz equations in the rest system $K$ that include convection currents (with sources): the component form of Faraday’s law of induction in Hertz's formalism:

\begin{equation} \label{AMM}
\begin{aligned}
\frac{1}{c}\frac{\partial L}{\partial t} &= \frac{\partial Y}{\partial z} - \frac{\partial Z}{\partial y}, \\
\frac{1}{c}\frac{\partial M}{\partial t} &= \frac{\partial Z}{\partial x} - \frac{\partial X}{\partial z}, \\
\frac{1}{c}\frac{\partial N}{\partial t} &= \frac{\partial X}{\partial y} - \frac{\partial Y}{\partial x},  \quad \textsc{Faraday’s law}
\end{aligned}
\end{equation}
and the component form of the Ampère–Maxwell law in the Hertz formalism:

\begin{equation} \label{AM}
\begin{aligned}
\frac{1}{c}\left(u_x \rho + \frac{\partial X}{\partial t}\right) &= \frac{\partial N}{\partial y} - \frac{\partial M}{\partial z}, \\
\frac{1}{c}\left(u_y \rho + \frac{\partial Y}{\partial t}\right) &= \frac{\partial L}{\partial z} - \frac{\partial N}{\partial x}, \\
\frac{1}{c}\left(u_z \rho + \frac{\partial Z}{\partial t}\right) &= \frac{\partial M}{\partial x} - \frac{\partial L}{\partial y},  \quad \textsc{Ampère–Maxwell}
\end{aligned}
\end{equation}
[$(u_x,u_y,u_z)$ denotes the velocity of the charge] together with Gauss’s law for the divergence of the electric field:%
\footnote{In Einstein’s 1905 notation, following the Maxwell–Hertz formalism, Gauss’s law is written simply as equation \eqref{GS}, with no factor of $4\pi$. This reflects the cgs/Hertz choice of units, in which the 
electrical field components $X,Y,Z$ and the charge density $\rho$ are defined so that the usual Gaussian factor $4\pi$ does not appear. In Gaussian units one would have: $\nabla \cdot \mathbf{E} = 4\pi \rho$ and in SI, $\nabla \cdot \mathbf{E} = \rho/\varepsilon_0$.}

\begin{equation} \label{GS}
\rho = \frac{\partial X}{\partial x} + \frac{\partial Y}{\partial y} + \frac{\partial Z}{\partial z}.    
\end{equation}

\item Applying:
I. the \emph{Lorentz transformation} \eqref{eq12} of section \S 3, i.e., the derivatives of the Lorentz transformation,%
\footnote{Deriving derivatives of the Lorentz transformation: First, we express $\frac{\partial}{\partial t}$ in terms of derivatives with respect to $\tau, \xi, \eta, \zeta$. To do this, we apply the chain rule. 
For the operator $\frac{\partial}{\partial t}$ transforms as:
\begin{equation} \label{chain-1}
\frac{\partial}{\partial t}
= \frac{\partial \tau}{\partial t}\frac{\partial}{\partial \tau}
+ \frac{\partial \xi}{\partial t}\frac{\partial}{\partial \xi}
+ \frac{\partial \eta}{\partial t}\frac{\partial}{\partial \eta}
+ \frac{\partial \zeta}{\partial t}\frac{\partial}{\partial \zeta}.
\end{equation}
Similarly, the operator $\frac{\partial}{\partial x}$ transforms as:
\begin{equation} \label{chain-2}
\frac{\partial}{\partial x}
= \frac{\partial \tau}{\partial x}\frac{\partial}{\partial \tau}
+ \frac{\partial \xi}{\partial x}\frac{\partial}{\partial \xi}
+ \frac{\partial \eta}{\partial x}\frac{\partial}{\partial \eta}
+ \frac{\partial \zeta}{\partial x}\frac{\partial}{\partial \zeta}.
\end{equation}
Since $\eta = y$ and $\zeta = z$ do not depend on $t$ or $x$, the corresponding derivatives vanish in both cases.
From the definitions of the Lorentz transformation \eqref{eq12}:

\begin{equation}
\frac{\partial \tau}{\partial t} = \gamma, \quad
\frac{\partial \xi}{\partial t} = -\gamma v, \quad
\frac{\partial \eta}{\partial t} = 0, \quad
\frac{\partial \zeta}{\partial t} = 0. 
\end{equation}
So plugging this into the chain rule \eqref{chain-1}, we get:
\begin{equation} 
\frac{\partial}{\partial t}
= \gamma \frac{\partial}{\partial \tau}
- \gamma v \frac{\partial}{\partial \xi}.    
\end{equation}
We do the same for $\frac{\partial}{\partial x}$. 
From the definitions of the Lorentz transformation \eqref{eq12}:

\begin{equation}
\frac{\partial \tau}{\partial x} = -\gamma \frac{v}{c^{2}}, \quad
\frac{\partial \xi}{\partial x} = \gamma. 
\end{equation}
So plugging this into the chain rule \eqref{chain-2}, we get:

\begin{equation} 
\frac{\partial}{\partial x}
= -\gamma \frac{v}{c^{2}} \frac{\partial}{\partial \tau}
+ \gamma \frac{\partial}{\partial \xi}. 
\end{equation}
For $y, z$, since $\eta = y$ and $\zeta = z$: 
\begin{equation} 
\frac{\partial}{\partial y} = \frac{\partial}{\partial \eta},
\quad
\frac{\partial}{\partial z} = \frac{\partial}{\partial \zeta}.
\end{equation}
Thus, the derivatives of the Lorentz transformation \eqref{eq12} transform as \eqref{12-2}.} 

\begin{equation} \label{12-2}
\begin{aligned}
\frac{\partial}{\partial t} &= \gamma \left(\frac{\partial}{\partial \tau} - v \frac{\partial}{\partial \xi} \right), \\ 
\frac{\partial}{\partial x} &= \gamma \left( \frac{\partial}{\partial \xi} - \frac{v}{c^{2}} \frac{\partial}{\partial \tau} \right), \\ 
\frac{\partial}{\partial y} &= \frac{\partial}{\partial \eta}, \quad \frac{\partial}{\partial z} = \frac{\partial}{\partial \zeta};
\end{aligned}
\end{equation}

II. the \emph{electromagnetic transformation} for the field components \eqref{EM}, and: 

III. the inverse field transformations \eqref{EM-1} of section \S 6, 
to the derivatives and fields $(X,Y,Z)$, Einstein obtains equations of the same functional form that hold in system $k$.

We start from the $x$–component of the Ampère–Maxwell equation in $K$ \eqref{AM}:
\begin{equation} \label{AMP}
\frac{1}{c}\!\left(\rho\,u_x + \frac{\partial X}{\partial t}\right)
= \frac{\partial N}{\partial y} - \frac{\partial M}{\partial z}.
\end{equation}
We apply the derivative rules (\emph{Lorentz transformation}) \eqref{12-2} and the \emph{field transformation} \eqref{EM} together with the inverse transformation \eqref{EM-1}. We also transform the current $\mathbf{j}=\rho\,\mathbf{u}$ term on the left-hand side of the Ampère–Maxwell equation \eqref{AMP}. We carry all this through, and we finally land on the transformed Ampère–Maxwell equation \eqref{NM}.

Specifically, using equations \eqref{12-2} and \eqref{EM-1}, we first transform the right-hand side (the curl terms) of the Ampère–Maxwell equation \eqref{AMP}: 

\begin{equation} \label{R}
\begin{aligned}
\frac{\partial N}{\partial y} - \frac{\partial M}{\partial z}
&= \frac{\partial}{\partial \eta}\Big[\gamma\big(N' + \tfrac{v}{c} Y'\big)\Big]
 - \frac{\partial}{\partial \zeta}\Big[\gamma\big(M' - \tfrac{v}{c} Z'\big)\Big] \\
&= \gamma\left(\frac{\partial N'}{\partial \eta} - \frac{\partial M'}{\partial \zeta}\right)
 + \gamma \frac{v}{c}\left(\frac{\partial Y'}{\partial \eta} + \frac{\partial Z'}{\partial \zeta}\right).
\end{aligned}
\end{equation}
Next, we transform the time derivative on the left-hand side of equation \eqref{AMP}. From the derivative rules (\emph{Lorentz transformation}) \eqref{12-2} and $X=X'$: 
\begin{equation} \label{T}
\frac{\partial X}{\partial t}
= \gamma \left(\frac{\partial X'}{\partial \tau} - v \frac{\partial X'}{\partial \xi}\right).
\end{equation}

\item We invoke the \emph{principle of relativity}, and require that the Ampère–Maxwell equations in $k$ have the same functional form as equation \eqref{AM} in $K$. This requires an invariant source transformation; specifically, we need the \emph{charge-density transformation}. 

\noindent We consider the two systems of coordinates, $K$ and $k$.  
The system $k$ moves with velocity $v$ in the direction of the positive $x$–axis 
relative to the system $K$.
The "density of electricity" (charge density) shall be $\rho$, and the convection current is:

\begin{equation}
\rho \mathbf{u} = \bigl(\rho u_x, \; \rho u_y, \; \rho u_z \bigr).    
\end{equation}
From the Maxwell–Hertz equations with convection current, one obtains the 
\emph{continuity equation} (local conservation of charge) for the density of electricity $\rho$ and the convection current $\,\mathbf{j}=\rho \mathbf{u}\,$:%
\footnote{Gauss’s law \eqref{GS} defines charge density $\rho$ as the divergence of the electric field $(X, Y, Z)$. The continuity equation then enforces that this charge density $\rho$, together with current $\rho \mathbf{u}$, obeys local conservation.}
\begin{equation} \label{eq12-1}
\frac{\partial \rho}{\partial t} 
+ \frac{\partial (\rho u_x)}{\partial x} 
+ \frac{\partial (\rho u_y)}{\partial y} 
+ \frac{\partial (\rho u_z)}{\partial z} = 0.
\end{equation}
If we substitute the derivatives of the Lorentz transformation \eqref{12-2} into the continuity equation \eqref{eq12-1}, we obtain:

\begin{align} \label{eq12-9}
0 &= \underbrace{\gamma\!\left(\frac{\partial}{\partial \tau}-v\frac{\partial}{\partial \xi}\right)\!\rho}_{\displaystyle \frac{\partial \rho}{\partial t}}
 \;+\; \underbrace{\gamma\!\left(\frac{\partial}{\partial \xi}-\frac{v}{c^{2}}\frac{\partial}{\partial \tau}\right)\!(\rho u_x)}_{\displaystyle \frac{\partial(\rho u_x)}{\partial x}}
 \;+\; \underbrace{\frac{\partial}{\partial \eta}(\rho u_y)}_{\displaystyle \frac{\partial(\rho u_y)}{\partial y}}
 \;+\; \underbrace{\frac{\partial}{\partial \zeta}(\rho u_z)}_{\displaystyle \frac{\partial(\rho u_z)}{\partial z}} \\[6pt]
&= \gamma\frac{\partial\rho}{\partial \tau}
   -\gamma v\frac{\partial\rho}{\partial \xi}
   +\gamma\frac{\partial(\rho u_x)}{\partial \xi}
   -\gamma \frac{v}{c^{2}}\frac{\partial(\rho u_x)}{\partial \tau}
   +\frac{\partial(\rho u_y)}{\partial \eta}
   +\frac{\partial(\rho u_z)}{\partial \zeta}.
\end{align}
Now we group the $\partial/\partial\tau$ terms and the $\partial/\partial\xi$ terms (noting that
$v$, $c$, and hence $\gamma$ are constants):

\begin{align}
0
&= \frac{\partial}{\partial \tau}\!\left[\gamma\rho
     -\gamma\frac{v}{c^{2}}(\rho u_x)\right]
   \;+\; \frac{\partial}{\partial \xi}\!\left[\gamma(\rho u_x)
     -\gamma v\,\rho\right]
   \;+\; \frac{\partial(\rho u_y)}{\partial \eta}
   \;+\; \frac{\partial(\rho u_z)}{\partial \zeta}.
\end{align}
Equivalently:
\begin{equation}
\frac{\partial}{\partial \tau}\,\gamma\!\left(\rho-\frac{v}{c^{2}}\rho u_x\right)
+\frac{\partial}{\partial \xi}\,\gamma\!\left(\rho u_x - v\rho\right)
+\frac{\partial}{\partial \eta}\,(\rho u_y)
+\frac{\partial}{\partial \zeta}\,(\rho u_z)=0.
\end{equation}
It therefore follows that in the system $k$, the following quantities are to be taken:%
\footnote{Einstein explicitly writes out only the first transformed equation.}
\begin{equation} \label{eq12-6}
\boxed{\rho'= \gamma \left(\rho - \frac{v}{c^{2}} \rho u_x\right)},  \quad  \rho' u_\xi =\gamma (\rho u_x - v \rho), \quad \rho' u_\eta = \rho u_y, \quad \rho' u_\zeta = \rho u_z.
\end{equation}
Assuming $\rho\neq 0$, we write for the $x$-component:
\begin{equation} \label{eq12-3}
u_\xi=\frac{\rho' u_\xi}{\rho'} =\frac{\gamma(\rho u_x - v\rho)}{\gamma\!\left(\rho-\dfrac{v}{c^{2}}\rho u_x\right)} = \frac{\rho(u_x - v)}{\rho\!\left(1-\dfrac{v}{c^{2}}u_x\right)}  =  \frac{u_x - v}{1-\dfrac{v}{c^{2}}u_x}\, .
\end{equation}
For the $y$-component:
\begin{equation} \label{eq12-4}
u_\eta = \frac{\rho' u_\eta}{\rho'} =\frac{\rho u_y}{\gamma\!\left(\rho-\dfrac{v}{c^{2}}\rho u_x\right)} = \frac{\rho u_y}{\gamma \rho\!\left(1-\dfrac{v}{c^{2}}u_x\right)}=\frac{u_y}{\gamma\!\left(1-\dfrac{v}{c^{2}}u_x\right)}\, .   
\end{equation}
For the $z$-component:
\begin{equation} \label{eq12-5}
u_\zeta =\frac{\rho' u_\zeta}{\rho'} = \frac{\rho u_z}{ \gamma \rho\!\left(1-\dfrac{v}{c^{2}}u_x\right)}=\frac{u_z}{\gamma\!\left(1-\dfrac{v}{c^{2}}u_x\right)}. 
\end{equation}
Finally, once working out each coordinate separately --- equations \eqref{eq12-3}, \eqref{eq12-4}, and \eqref{eq12-5} --- we gather them into the full set of relativistic velocity transformations \cite{Einstein05}: 
\begin{equation} \label{eq96}
\boxed{u_{\xi}= \frac{u_x - v}{1 - \tfrac{u_x v}{c^2}}, \quad  
u_{\eta} = \frac{u_y}{\gamma \left( 1 - \tfrac{u_x v}{c^2} \right)}, \quad
u_{\zeta} = \frac{u_z}{\gamma \left( 1 - \tfrac{u_x v}{c^2} \right)},}
\end{equation}

\item Now we put \eqref{R}, \eqref{T}, and \eqref{eq12-6} into the Ampère–Maxwell equation \eqref{AMP}. From the Gauss law \eqref{GS} in system $K$, using the derivative rules (\emph{Lorentz transformation}) \eqref{12-2} and the inverse transformation \eqref{EM-1}, one finds the transformed Gauss law in system $k$ \cite{Einstein05}:
\begin{equation}
\rho' = \frac{\partial X'}{\partial \xi}
      + \frac{\partial Y'}{\partial \eta}
      + \frac{\partial Z'}{\partial \zeta},
\end{equation}
and the transformed Ampère–Maxwell equation in system $k$ \cite{Einstein05}:

\begin{equation} \label{NM}
\frac{1}{c}\left(\rho' u_\xi + \frac{\partial X'}{\partial \tau}\right)
= \frac{\partial N'}{\partial \eta} - \frac{\partial M'}{\partial \zeta}.
\end{equation}

\item The transformed velocity vector $(u_\xi,\,u_\eta,\,u_\zeta)$ \eqref{eq96} coincides exactly with the result obtained from the velocity–addition law in section \S 5. 

If $k$ moves with velocity $v$ along the $x$--axis of $K$, then the transformation laws are given by equation \eqref{eq96}.
Einstein’s section \S 5 expresses the velocities in $K$ in terms of those in $k$, so we invert the formulas.

1. \emph{The Longitudinal component}:
Starting with: 

\begin{equation}
u_{\xi}= \frac{u_x - v}{1 - \tfrac{u_x v}{c^2}},        
\end{equation}
we expand the left-hand side:

\begin{equation}
u_{\xi}\!\left(1-\frac{u_x v}{c^2}\right) =  u_{\xi} - \frac{u_{\xi} u_x v}{c^2} = u_x - v.
\end{equation}
We collect terms in $u_x$; bring the terms with $u_x$ to one side, constants to the other:
\begin{equation}
u_{\xi} + v = u_x + \frac{u_{\xi} u_x v}{c^2}.
\end{equation}
Factor $u_x$ on the right-hand side:
\begin{equation}
u_{\xi} + v = u_x \left(1 + \frac{u_{\xi} v}{c^2}\right).
\end{equation}
Finally, we get \cite{Einstein05}:
\begin{equation} \label{ux}
\boxed{u_x=\frac{u_{\xi}+v}{1+\tfrac{u_{\xi} v}{c^2}}}. 
\end{equation}

2. \emph{Transverse components}:
From the general law \eqref{eq96}:
\begin{equation} \label{uyy}
u_{\eta}=\frac{u_y}{\gamma\left(1-\tfrac{u_x v}{c^2}\right)}, \qquad \text{we get:} \quad u_y=u_{\eta}\,\gamma\!\left(1-\tfrac{u_x v}{c^2}\right).
\end{equation}
Substituting $u_x$ found in equation \eqref{ux} into equation \eqref{uyy}, yields:

\begin{equation}
\begin{aligned}  
&1-\frac{u_x v}{c^2}
=1-\frac{v}{c^2}\,\frac{u_{\xi}+v}{1+\tfrac{u_{\xi} v}{c^2}} 
=\frac{1+\tfrac{u_{\xi} v}{c^2}-\tfrac{u_{\xi} v}{c^2}-\tfrac{v^2}{c^2}}{1+\tfrac{u_{\xi} v}{c^2}} \\
&=\frac{1-\tfrac{v^2}{c^2}}{1+\tfrac{u_{\xi} v}{c^2}} 
=\frac{1/\gamma^{2}}{1+\tfrac{u_{\xi} v}{c^2}}.  \quad \text{Thus:}
\end{aligned}
\end{equation}

\begin{equation} \label{uy}
u_y
=u_{\eta}\,\gamma \cdot \frac{1/\gamma^{2}}{1+\tfrac{u_{\xi} v}{c^2}}
=\frac{u_{\eta}\,\sqrt{1-\tfrac{v^{2}}{c^{2}}}}{1+\tfrac{u_{\xi} v}{c^2}}.
\end{equation}
By the same reasoning:
\begin{equation} \label{uz}
u_z
=\frac{u_{\zeta}\,\sqrt{1-\tfrac{v^{2}}{c^{2}}}}{1+\tfrac{u_{\xi} v}{c^2}}.
\end{equation}

3. \emph{Einstein’s special case}:
In section \S 5, Einstein considers uniform motion in $k$ \cite{Einstein05}:
\begin{equation}
\xi=w_{\xi}\tau, \quad \eta=w_{\eta}\tau, \quad \zeta=0, \quad \text{so that:}
\end{equation}

\begin{equation}
u_{\xi}=w_{\xi}, \qquad u_{\eta}=w_{\eta}, \qquad u_{\zeta}=0.
\end{equation}
Plugging this into the formulas \eqref{ux}, \eqref{uy} and \eqref{uz} yields the motion in $K$:
\begin{equation}
u_x=\frac{w_{\xi}+v}{1+\tfrac{v w_{\xi}}{c^2}}, \qquad
u_y=\frac{w_{\eta}\sqrt{1-\tfrac{v^{2}}{c^{2}}}}{1+\tfrac{v w_{\xi}}{c^2}}, \qquad
u_z=0.
\end{equation}
Since the velocities are constants, the particle has uniform motion in $K$. Integrating gives the same linear relations Einstein prints in section \S 5 of his paper \cite{Einstein05}:
\begin{equation} \label{E}
\boxed{x=\frac{w_{\xi}+v}{1+\tfrac{v w_{\xi}}{c^2}}\,t, \qquad
y=\frac{\sqrt{1-\tfrac{v^{2}}{c^{2}}}}{1+\tfrac{v w_{\xi}}{c^2}}\,w_{\eta}\,t, \qquad
z=0.}
\end{equation}

Thus, Einstein’s printed derivation in section \S 5 follows the same method as the derivation that leads to the full three-dimensional law \eqref{eq96}. Still, he demonstrates it only in a special case, namely the two-dimensional case with $u_{\zeta} = 0$ (no $z$ component).

Now, Einstein closes the loop in section \S 9: "Since — as follows from the theorem of addition of velocities (\S 5) the velocity vector $(u_\xi, u_\eta, u_\zeta)$ [equation \eqref{eq96}] is nothing other than the velocity of the electric masses as measured in the system $K$, it is thereby shown that, based on our kinematical principles, the electrodynamical foundation of Lorentz’s theory of the electrodynamics of moving bodies corresponds to the principle of relativity"  \cite{Einstein05}.  
\end{enumerate}

\subsection{The Addition Law that Made the Difference} \label{Ad}

Ginoux observes that in \S 5 of his 1905 paper, "Einstein establishes the new relativistic velocity addition law. 
The simplest way to do this is to write the ratio of the total differentials of the first two equations of the [Lorentz] transformation... 
Curiously, this is not the way (used by Poincaré in his memoir to the \emph{Rendiconti}) that Einstein will proceed" \cite{Gin}. 
Ginoux is correct: Einstein did not, and in 1905 could not, adopt this purely formal route, since his method was heuristic in character.\footnote{For a detailed reconstruction of Einstein’s original derivation of the velocity--addition law, see my recent paper \cite{Weinstein}.}  

Einstein and Poincaré run in \emph{complementary} directions toward the same invariant object. In modern language, both demonstrate that the four-current $(c \rho,\, \rho u_x,\, \rho u_y,\, \rho u_z)$ must transform as a four-velocity vector $\gamma(c, u_x, u_y, u_z)$, so that the continuity equation \eqref{eq12-1} (Einstein) and \eqref{eqC-2}
(Poincaré) remains form-invariant. However, they navigate that terrain from different entry points. Poincaré first applies the Lorentz transformation \eqref{eq97} and differentiates it to obtain the velocity transformation \eqref{eq101}. He then inserts those velocity relations into the transformation of the current-density variables \eqref{eqtr} and verifies that the continuity equation \eqref{eqC-2} retains its form. Einstein demands that the inhomogeneous Maxwell--Hertz equations with sources \eqref{AMM} and \eqref{AM} keep their form between systems $K$ and $k$. Then he uses the continuity
equation \eqref{eq12-1} together with Lorentz-transformed derivatives and fields to read off the source transformation \eqref{eq12-6}, i.e., how $\rho$ and $(\rho u_x,\rho u_y,\rho u_z)$ transform. Finally, he takes ratios $u_i'=\tfrac{\rho' u'_i}{\rho'}$ to obtain the velocity transformation \eqref{eq96} (see step 3 in Einstein's derivation).

However, the crucial point comes at the end of section \S 9 of Einstein's paper. In section \S 5, Einstein had already derived the relativistic velocity--addition law \eqref{eq41} as a purely kinematical consequence of the Lorentz transformation. 
Now, in \S 9, the same law \eqref{eq96} reappears in electrodynamics, as the transformation of the convection current in the inhomogeneous Maxwell--Hertz equations with sources \eqref{AMM}, \eqref{AM}. By comparing the kinematical result of \S 5 with the dynamical result of \S 9, Einstein shows that the two coincide. Thus, the kinematical principles of relativity and the constancy of the velocity of light provide a foundation on which the electrodynamics of moving bodies itself rests. 
The electrodynamic basis of Lorentz's theory is thereby brought into exact agreement with the relativity principle \cite{Einstein05}. This closure of the loop between kinematics and dynamics constitutes the true revolution, which is not found in Poincaré. Poincaré’s derivation---differentiating the Lorentz transformation and dividing differentials---is a formal calculation embedded in Lorentz’s electron theory.

Thus, Einstein and Poincaré intersect mathematically, but there is an abyss between them conceptually. Poincaré’s procedure refines Lorentz’s electron
theory by means of formal differentiation and verification of charge conservation, while Einstein’s procedure reconstructs the foundations of kinematics itself. What appears as the same velocity and charge
transformations in both derivations thus conceal two profoundly different programs: one entrenched in Lorentz’s dynamics, the other grounded in
Einstein’s new relativistic kinematics. 

\section{In the Shadow of Light Beams}

\subsection{The Relativity of Recognition}

Ginoux undertakes a comparative study of how special relativity entered the scientific cultures of France, Germany, the Netherlands, the United Kingdom, and Italy. His country-by-country autopsy reframes the “discovery” of special relativity as something decided less by priority claims than by the patterns of reception. In France, Germany, the Netherlands, the United Kingdom, and Italy, he tallies who cited whom—and, just as revealingly, who did not—to show how credit consolidated around “Einstein,” while Poincaré’s very real contributions (the relativity postulate, group structure of the Lorentz transformations, velocity addition, invariant and least-action principles, even early gravitational remarks) receded to the footnotes \cite{Gin}. The explicit narrative is curricular and editorial uptake; the tacit one is the sorting power of gatekeeping, branding, and the authority vested in certain styles of physics and lines of mentorship.

In France, Langevin—well aware of Poincaré’s 1905 note and 1906 Palermo memoir—nonetheless organized his influential lectures so that generations of students encountered the subject under the “Einstein” label. In Germany, Planck’s early public allegiance to “die Einsteinsche Relativitätstheorie,” amplified by his seminars and doctoral training, and Minkowski’s four-dimensional reformulation, cemented an Einstein–Minkowski partnership in which Poincaré’s mathematics was freely used but rarely mentioned. In Leiden, Lorentz oscillated: admiring Poincaré’s derivations in private, yet in public lectures emphasizing Einstein, before later striking a compromise that largely stabilized the Einstein-first storyline. In Britain, Joseph Larmor and his student Ebenezer Cunningham taught the Lorentz-Einstein relativity from 1907, with Poincaré’s name surfacing only belatedly in Cunningham’s 1914 book. Italy was a partial exception: mathematicians like Levi-Civita and Volterra kept Poincaré’s \emph{Rendiconti} memoir visible. However, the broader teaching frame still spoke of a Lorentz–Einstein synthesis \cite{Gin}.

For Ginoux, the lesson is that these convergences were hardly accidental because: 

1. \emph{Venue mattered}: Poincaré’s memoir, tucked away in a mathematics journal, did not compete with Einstein’s 1905 paper in \emph{Annalen der Physik}. 

2. \emph{Style mattered}: Poincaré’s abstract, group-theoretic exposition, still shading toward the ether, contrasted with Einstein’s stripped-down operational kinematics of clocks, rods, and thought experiments. 

3. \emph{Networks mattered}: Planck’s editorship and seminars, Langevin’s lectures, Minkowski’s geometry. 

4. \emph{Timing and narrative economy mattered}: the Kaufmann controversy and Bucherer’s 1908 reprieve, Poincaré’s death in 1912, and finally the 1915 triumph of general relativity. 

The pattern, Ginoux concludes, is clear enough: in the very places where relativity became a living subject of teaching and research, Poincaré was acknowledged but set aside, while “Einstein” supplied the name under which the theory entered physics culture—and once established there, the label required little further defense \cite{Gin}.

Ginoux’s examination of the Nobel dossiers casts the episode less as a mystery of merit than as a study in how institutions quietly decide what will be remembered. Despite repeated nominations, Poincaré’s case never advanced. The committee preferred tangible experiment to abstract theory, mistrusted mathematical style in physics, and managed national sensibilities with a cautious hand. The reports are respectful, sometimes admiring, but never decisive. His work is acknowledged in passing, yet consistently absorbed into “Lorentz’s theory” or filed under collective progress. By the time of his death in 1912, the record had settled into a pattern of praise without credit—an archive of tributes that effectively erased their subject.

Einstein’s file, by contrast, circulated under a single name and gathered momentum even before the 1919 eclipse observations. Where Poincaré was always present yet never quite nominated in his own right, Einstein appeared as the unmistakable author of a new physics awaiting only experimental vindication. The symmetry is telling: in lectures, textbooks, and journals, relativity came to be taught under one label; in the Nobel archive, recognition followed the same logic. If the Prize was meant to adjudicate scientific achievement, here it functioned more as an instrument of inscription—leaving Poincaré remembered by silence, and Einstein remembered by name \cite{Gin}.

Ginoux’s survey traces the mechanics of consecration. Poincaré’s mathematics circulated, but it was “Einstein” that the key journals and lecture halls chose to canonize, the name under which relativity became teachable. Priority was not overturned so much as edited out; silence, placement, and curricular economy determined who was remembered and who was background. Yet even Einstein, the supposed beneficiary of this process, was not immune to its reversals. As Abraham Pais notes, when the 1919 eclipse results vaulted him to international renown, the Nobel Committee still withheld acknowledgment of relativity; in 1921, he was awarded the prize with the explicit proviso that relativity was excluded \cite{Pa1, Pa2}. In Ginoux’s reconstruction, the moral is that credit can be erased; in Pais’s, that recognition may come only once safely detached from the work itself. The paradox is plain. Poincaré disappeared through omission, while Einstein prevailed through branding, and both illustrate that, in the end, it was not the theory but the machinery of recognition that ultimately decided how relativity would be remembered.

\subsection{\emph{La problématique}: Why Did Poincaré Not Claim Authorship of Special Relativity?}

Ginoux points to Poincaré’s own Nobel application, in which he described his 1905 \emph{Rendiconti di Palermo} memoir not as a theory of relativity, but as a polishing of Lorentz’s electron dynamics. In a memorandum to Gaston Darboux, Poincaré listed thirteen areas of his scientific work. The ninth reads with a curious modesty: 
“I have published an article in \emph{Rendiconti} in which I explain the theory of Lorentz on the Dynamics of the Electron and in which I believe I have succeeded in removing the last difficulties and giving it perfect coherence” \cite{Gin}.

The irony is plain. Here was the same Poincaré whose mathematical vision cleared the path for relativity, presenting himself not as the founder of a new theory but as Lorentz’s faithful expositor. And yet, a few years later, in his 1912 London lecture, he was speaking the idiom of Minkowski space and edging toward Einstein’s own perspective. 

In his London lecture of May 1912, Poincaré spoke of a “révolution en physique.” No ether was invoked. Instead, he described how a sphere for one observer becomes an ellipsoid for another, how simultaneity dissolves once observers move with respect to each other, and how time itself interlaces with space as a fourth dimension—mathematically rotated by Lorentz transformations. One even hears the unmistakable echo of Minkowski in his remark that the fourth coordinate is best taken as $t\sqrt{-1}$. Finally, he underlined that in this nouvelle mécanique, causal influence can propagate no faster than light, so that certain pairs of events can stand in no relation of cause and effect at all \cite{Poi12}.

Read in isolation, the passage might almost pass for a lecture by Einstein himself—save for the French accent and the courtesy of quotation marks. By then, however, Einstein had already made simultaneity operational, abandoned the ether, and embraced Minkowski’s geometry. The irony is hard to miss: Ginoux asks why Poincaré never claimed authorship of relativity \cite{Gin}, yet Poincaré’s own words, in the last public address of his life, sound less like a claimant and more like a convert. The revolution, it seems, was already underway—only the naming rights remained unspoken.

Even Bergson would later recall that it was Paul Langevin’s Bologna talk of 1911 that first drew his attention to Einstein’s ideas \cite{Ber}. If so, the Parisian philosophers discovered relativity at second hand—through Langevin—while Poincaré himself, in his final year, was already speaking as though he had joined the chorus. One might almost say that France met Einstein not face to face, but in translation.

\section{Conclusion}

Ginoux carefully tallies the calendar. Poincaré’s memoir between 27 April and 5 June 1905, his note on 5 June; Einstein’s article written between 18 May and 30 June: "For Poincaré, it was established in Chap. 5 that he had written his memoir to the \emph{Rendiconti} between the 27th April and the 5th June 1905 and that his note to the C.R.A.S. presented on 5 June 1905 was only a summary of this memoir. For Einstein, it was established in Chap. 7 that he had written his article published in the \emph{Annalen der Physik} between the 18th May and the 30th June 1905." 
From this, he concludes that special relativity must have one father: "if we accept the indisputable anteriority of Poincaré’s contributions, then 
there is only one real question that needs to be answered if we are to put a definitive end to this controversy: Why didn’t Poincaré claim the special theory of relativity?" \cite{Gin}. But calendar arithmetic cannot settle paternity. 

Ginoux assembles a long catalogue of Poincaré’s remarks on the principle of relativity, simultaneity, synchronization, and even the ether, as if their very abundance proved equivalence with Einstein. He lines up algebraic formulas until Einstein and Poincaré look interchangeable, then suggests Einstein only took the tedious route to what Poincaré had already simplified \cite{Gin}. 

Yet the resemblance of formulas should not obscure the difference of foundations.
The decisive difference lies not in whether both men spoke of rods, clocks, or signals, but in what they did with them. Poincaré preserved the ether, corrected Lorentz's electron theory within electrodynamics, and treated local time as a useful fiction; Einstein discarded the ether, made simultaneity operational, and showed that the space and time transformations express a new conception of space and time. That is why time dilation, the relativity of simultaneity, and the invariance of $c$ emerge naturally in Einstein’s paper but not in Poincaré’s memoir. Algebraic coincidence is not conceptual equivalence.

Blurring the line so that similar formulas obscure foundational differences points to a larger caution in historical analysis: priority accounts risk distortion when the first appearance of a formula is taken as the origin of a theory. Formal precedence matters, but explaining why Einstein’s version became canonical requires attention to conceptual framing, pedagogical dissemination, and the redefinition of physical meaning. Concordant equations do not by themselves establish theoretical identity. The Lorentz transformation could serve different research programs, and it was precisely the divergence—constructive, ether-based dynamics versus principle-driven kinematics that determined their distinct historical trajectories.  

One suspects that, had he been listening, Einstein would have let the discussion run its course before offering only a shrug and the reminder that equations, like jokes, are all in the telling.

\section*{Acknowledgment}

I want to thank the late Prof. John Stachel for spending numerous hours with me discussing Einstein's relativity at the Center for Einstein Studies at Boston University.

\end{document}